\documentclass{aa}

\usepackage{natbib}
\usepackage{graphicx}
\usepackage[mathcal]{eucal}
\usepackage{amssymb} 
\usepackage{revsymb}	
\usepackage{amsmath}
\usepackage[usenames]{color}	
\usepackage{epstopdf}	
\long\def\jumpover#1{{}}

\newcommand{\derivp} [2] {\frac {\partial #1 } {\partial #2} }

\newcommand{\eq}[1] {Eq.~(\ref{#1})}
\newcommand{\eqs}[1] {Eqs.~(\ref{#1})}

\newcommand{\eqn} [1] {
\begin{equation}#1
\end{equation}}

\usepackage[applemac]{inputenc}

\def\acenA{$\alpha$~Cen~A}

\begin{document}

\defcitealias{Samadi09a}{Paper~I} 	

\title{The CoRoT\thanks{The CoRoT space mission, launched on
    December 27 2006, has been developped and is operated by CNES,
    with the contribution of Austria, Belgium, Brasil, ESA, Germany
    and Spain.} target HD~49933:}
\subtitle{2 - Comparison of theoretical mode amplitudes with observations}
\author{ R. Samadi \inst{1}   
\and H.-G. Ludwig\inst{2}
\and K. Belkacem\inst{1,3}  
\and M.J. Goupil\inst{1}    
\and O. Benomar\inst{4}  
\and B. Mosser\inst{1}  
\and M.-A. Dupret\inst{1,3}  
\and F. Baudin\inst{4}  
\and T. Appourchaux\inst{4}  
\and E. Michel\inst{1}  }

\institute{
Observatoire de Paris, LESIA, CNRS UMR 8109, Universit\'e Pierre et Marie Curie, Universit\'e Denis Diderot, 5 pl. J. Janssen, F-92195 Meudon, France \and 
Observatoire de Paris, GEPI, CNRS UMR 8111, 5 pl. J. Janssen, F-92195 Meudon, France\and 
Institut d'Astrophysique et de G\'eophysique de l'Universit\'e de
Li\`ege, All\'e du 6 Ao\^{u}t 17 - B 4000 Li\`ege, Belgium \and 
Institut d'Astrophysique Spatiale, CNRS UMR 8617, Universit\'e Paris XI, 91405 Orsay, France.}


\mail{Reza.Samadi@obspm.fr}

\date{\today} 

\titlerunning{The CoRoT target HD~49933: 2 - Comparison with observations}

\abstract
{The seismic  data obtained by CoRoT for the star HD~49933 enable us for the first time 
to measure \emph{directly}  the amplitudes and linewidths of 
  solar-like oscillations for a star other than the Sun.
From those measurements it is possible, as was done for the Sun, to
 constrain models of the excitation of acoustic modes by turbulent convection. }
{We compare a stochastic excitation model described in Paper~I with the asteroseismology data for HD~49933,
a star that is rather metal poor and 
significantly hotter than the Sun.}
{Using the seismic determinations of the mode linewidths 
detected by CoRoT for HD~49933 and the theoretical mode excitation
rates computed in Paper~I for the specific case
of HD~49933,  we derive the expected surface velocity amplitudes of the acoustic modes
detected in HD~49933. Using a calibrated quasi-adiabatic approximation relating the mode amplitudes
in intensity to those in velocity, we derive the expected values of
the mode amplitude in intensity.  }
{ Except at rather high frequency, our amplitude calculations are within 1-$\sigma$  error bars of the mode surface velocity spectrum derived with the HARPS  spectrograph.  The same is found with respect to the mode amplitudes in intensity derived for HD~49933 from the CoRoT data. On the other hand, at high frequency ($\nu \gtrsim~1.9 $~mHz),  our
calculations depart significantly  from the CoRoT and HARPS measurements.  
We show that assuming a solar metal abundance rather than the actual
metal abundance of the star would result in a larger
discrepancy with the seismic data. Furthermore, we present calculations
which assume the ``new'' solar chemical mixture to be in better agreement with
the seismic data than those that assumed the ``old''  solar chemical mixture.}
{  These results validate in the case of a star significantly hotter than the Sun and $\alpha$~
Cen~A the main assumptions in the model of stochastic
excitation. However, the discrepancies seen at high
frequency highlight some deficiencies of the modelling, whose origin remains to be understood.
We also show that it is important to take  the surface metal
  abundance of the solar-like pulsators into account. }


\maketitle

\section{Introduction}

The amplitudes of solar-like oscillations result from a balance between excitation and damping. 
The mode linewidths are directly related to the mode damping rates. 
Once we can measure the mode linewidths, we can derive the theoretical
value of the mode amplitudes  from
theoretical calculations of the mode excitation rates, which in turn
can be compared to the available seismic constraints. This comparison allows us to test
the model of stochastic mode excitation investigated in a companion
paper \citep[][hereafter Paper~I]{Samadi09a}.

As shown in Paper~I, a moderate deficit of the surface metal abundance results in a
significant decrease of the mode driving by turbulent
convection. Indeed, by taking into account the measured iron-to-hydrogen abundance ([Fe/H]) of
HD~49993 ([Fe/H]$=-0.37$), we have derived the theoretical values of the 
mode excitation rates ${\cal P}$ expected for this star. 
The resulting value of ${\cal P}$ is found to be about two times smaller than for a model
with the same gravity and effective temperature, but with a solar metal
abundance (i.e. [Fe/H]$=0$). 

The star HD~49933 was first observed in Doppler velocity by
\citet{Mosser05} with the HARPS spectrograph. 
More recently, this star has been observed  twice by CoRoT. A first time this was done  continuously during about 61 days (initial run, IR) and a second time continuously during about 137 days (first long run in the center direction, LRc01).  
The combined seismic analysis of these data \citep{Benomar09b} has
provided the mode linewidths as well as the amplitudes of the modes in intensity.  
Then, using mode linewidths  obtained for  HD 49933 with the CoRoT
data  and the theoretical mode excitation
rates (obtained in Paper~I), we derive the expected values of the mode
surface \emph{velocity} amplitudes.
We next compare these values  with the mode velocity spectrum derived 
following \citet{Kjeldsen05} with seismic data  from the  HARPS spectrograph
\citep{Mosser05}.

Mode amplitudes in terms of \emph{luminosity} fluctuations  have also been derived from the
CoRoT data for 17 radial orders.  These data provide us with not only a
constraint on the maximum of the mode amplitude but also with the frequency
dependence. 
The relative luminosity amplitudes
$\delta L/L$ are linearly related to the velocity amplitudes. This ratio is determined by the solution of
the \emph{non-adiabatic} pulsation equations and is independent of the 
stochastic excitation model \citep[see][]{Houdek99}.
Such a non-adiabatic calculation requires us to take into account,
not only the radiative damping, but also the coupling between
the pulsation and the turbulent convection.  However, there are currently
very significant uncertainties concerning the modeling of this
coupling \citep[for a recent review see ][]{Houdek08}.  
We relate further for the sake of simplicity the mode luminosity
amplitudes to computed mode velocity amplitudes by assuming adiabatic oscillations as
\citet{Kjeldsen95}. Such a relation is calibrated in order to reproduce the helioseismic data.

The comparison between theoretical values of the mode
amplitudes (both in terms of surface velocity and intensity) 
constitutes a test of the stochastic excitation model  with a star
significantly different from the Sun and 
{\acenA}. In addition  it is also
possible to test the validity of the calibrated quasi-adiabatic relation, 
since both mode amplitudes, in terms of surface 
velocity and intensity, are available for this star,

This paper is organized as follows:
We describe in Sect.~\ref{velocity} the way mode amplitudes
   in terms of surface velocity $v_s$ are  derived from
  the theoretical  values of ${\cal P}$ and from the measured mode
  linewidths ($\Gamma$). Then, we compare the theoretical values of the
  mode surface velocity with the seismic constraint obtained from HARPS
  observations. 
We describe in Sect.~\ref{intensity} the way mode
amplitudes in terms of intensity fluctuations $\delta L/L$ are derived from
  theoretical  values of $v_s$ and compare  $\delta L/L$ with the
  seismic constraints obtained from the CoRoT observations.
Finally, Sects. \ref{Discussion} and \ref{Conclusion} are dedicated to
a discussion and conclusion respectively.

\section{Surface velocity mode amplitude}
\label{velocity}

\subsection{Derivation of the surface velocity mode amplitude}
\label{calculation_velocity}

The intrinsic rms mode surface velocity $v_s$  is related to the mode
exitation rate $\mathcal P(\nu) $ and the mode linewidth $\Gamma(\nu) $
according to \citep[see, e.g.,][]{Baudin05}:
\eqn{
v_s(r_h,\nu)  = \sqrt{\mathcal P \over{ 2 \pi \, {\cal M}_h \, \Gamma  } }
\label{v_s}
}
where ${\cal P}$ is the mode excitation rate derived as
described in Paper~I, $\Gamma$ is the mode full width at half maximum (in $\nu$), $\nu
= \omega_{\rm osc} / 2 \pi$ the mode frequency and   
${\mathcal M}_h$ is the mode mass defined as:
\eqn{
{\cal M}_h= { I \over { \xi_{\rm r}^2(r_h) } }
\label{modemass}
} 
where $I$ is the mode inertia (see Eq.~(2) of Paper~I), $\xi_{\rm r}$ the radial mode
eigendisplacement, $r_h \equiv R + h$ the layer  in the atmosphere where the
mode is measured in radial velocity,  $R$  the radius at the
photosphere (i.e. at $T=T_{\rm 
  eff}$) and $h$ the height above the photosphere.

In Sect.~\ref{comparison_veloctiy} we will compare  estimated values of $v_s$ with the seismic constraint
obtained by \citet{Mosser05} with the HARPS spectrograph. We therefore
need to estimate  $v_s$ at the layer $h$ where the HARPS spectrograph is the most
sensitive to the mode displacement.
As discussed by \citet{Samadi08}, the seismic measurements obtained with
HARPS spectrograph are likely to arise from the  optical depth
$\tau_{\rm~500~nm} \simeq 0.013$, which corresponds to the
 depth where the potassium (K) spectral line is formed.
We then compute the mode mass at the layer $h$ associated with the  optical depth
$\tau_{\rm~500~nm}$ \citep{JCD82b}. For the model with [Fe/H]$=0$ (resp. [Fe/H]$=-1$) this optical depth
corresponds to $h \simeq$ 390~km (resp. $h \simeq$ 350~km). 

For the mode linewidth $\Gamma$ we use the seismic measurement
obtained from the seismic analysis of the CoRoT data performed by \citet{Benomar09b}.
This seismic analysis combined the two CoRoT runs available for HD~49933.
Two different approaches were considered in this analysis: one based on the maximum likelihood
estimator and the second one using the Bayesian approach coupled with a Markov Chains Monte Carlo
algorithm. 
The Bayesian approach remains in general more reliable even in low signal-to-noise conditions.
Nevertheless, in terms of mode amplitudes, mode heights and mode linewidths, both methods agree within 1-$\sigma$.
We will  consider here the seismic parameters and associated error bars obtained on the basis of the Bayesian approach.

\subsection{Comparison with the HARPS measurements}
\label{comparison_veloctiy}

 The seismic analysis in velocity has been performed by
  \citet{Mosser05} using data from the HARPS spectrograph.  
  The quality of these data is too poor to perform a direct comparison
  between the observed spectrum and the calculated amplitude spectrum
  ($v_s$, \eq{v_s}).  Indeed, the observed spectrum is highly affected
  by the day aliases.  Furthermore, the quality of the data does not
  allow to isolate individual modes, in particular modes of
  a different angular degree ($\ell$).  A consequence is that energies 
  of modes which are close in frequency are mixed.  

\begin{figure}[ht]        
\begin{center}
        \resizebox{\hsize}{!}{\includegraphics  {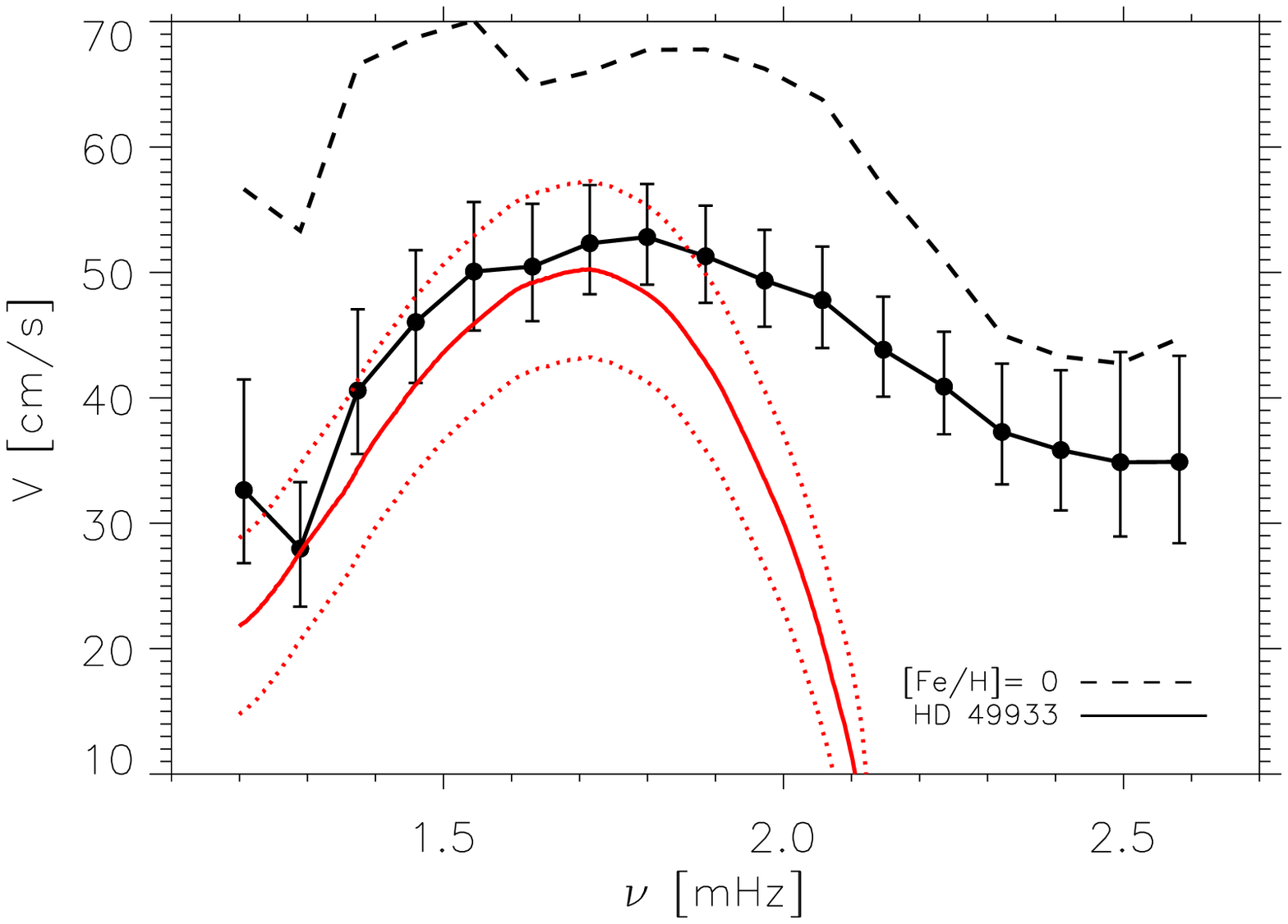}} 
         \resizebox{\hsize}{!}{\includegraphics  {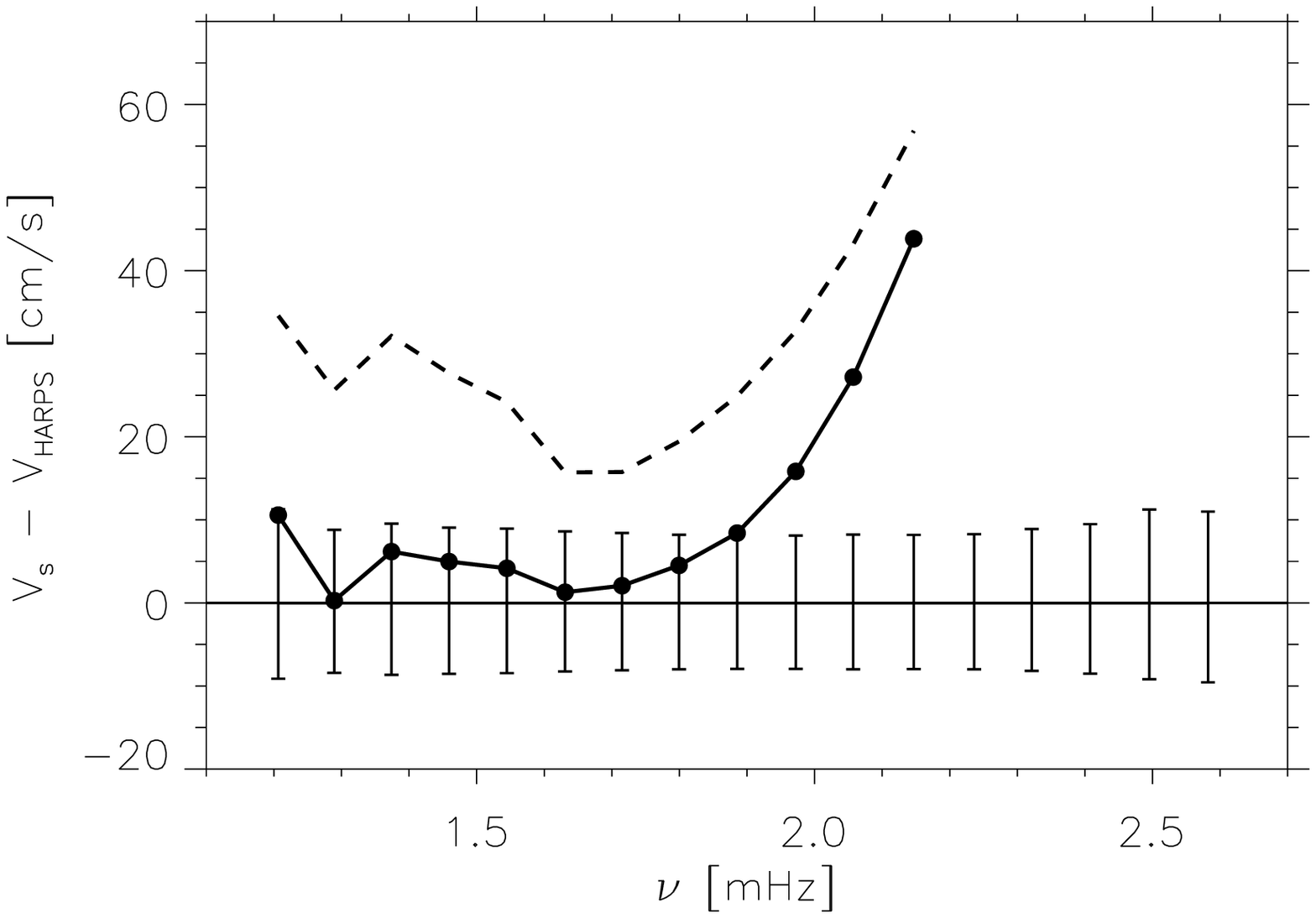}} 
       \end{center}    
        \caption{{\bf Top:}  Intrinsic mode surface velocity as a function of
  the mode frequency ($\nu$). The filled circles connected by the
  thick solid line correspond to the mode surface velocity ($v_s$) derived for
  HD~49933 according to Eq.~(\ref{v_s}), where the mode excitation rates ${\cal
  P}$ are derived as explained in Paper~I and the mode linewidths and
their associated error bars are derived by
\citet{Benomar09b} from the CoRoT data.
 The thick dashed   line corresponds to the mode velocity
 associated with the model with [Fe/H]$=0$. 
The thick and red solid line corresponds to the amplitude spectrum derived from the seismic observations obtained with the HARPS  spectrograph (see text). 
The dotted line corresponds to the 1-$\sigma$ domain associated with this measurement.
{\bf Bottom:} Differences between $v_s$ and
$v_{\rm HARPS}$. The 1-$\sigma$ error bars correspond to $\sigma_v
\equiv \sqrt{ \Delta   v_s^2 + \Delta v_{\rm HARPS}^2}$ (see text).}
        \label{velocity0}
        \end{figure}

  In order to measure the oscillation
  amplitude in a way that is independent of these effects, we have
  followed the method introduced by \citet[][see also \citet{Kjeldsen08}]{Kjeldsen05}.
  This method consists in deriving the oscillation amplitudes from the oscillation power density spectrum smoothed over typically four times the large separation (i.e. four radial orders). 
  Next, we multiply this smoothed spectrum by a coefficient in order to convert the \emph{apparent} amplitudes into \emph{intrinsic} amplitudes. 
  This coefficient takes into account the spatial response function of the angular degrees $\ell=$0, 1, 2 and 3 \citep[see][]{Kjeldsen08}. We have checked that the sensitivity of the visibility factor with the limb-darkening law is significantly smaller in comparison with the error associated  with the \citet{Mosser05} seismic measurements.   
  The amplitude spectrum $v_{\rm HARPS}$ derived following \citet{Kjeldsen05} is shown in Fig.~\ref{velocity0}. 
  The 1-$\sigma$ error bar associated with each values of $v_{\rm HARPS}$  is constant and equal to $\Delta V_{\rm  HARPS} = 7$~cm/s.

  The maximum of $v_{\rm HARPS}$  reaches $V_{\rm max}=
  50.2~\pm~7$~cm/s. By comparison, \citet{Mosser05} found a maximum
  of $40~\pm~10$~cm/s, which once converted into  \emph{intrinsic}
  amplitude represents a maximum of  $42~\pm~10$~cm/s. The difference between the two values is within the 1-$\sigma$ error bars. 
  The different value found by \citet{Mosser05} can be explained by the way the maximum of the mode amplitude was derived. 
  Indeed, \citet{Mosser05} have constructed synthetic time series based on a theoretical
low degree p-modes eigenfrequency pattern and theoretical mode lines widths \citep{Houdek99}. 
 The maximum  amplitudes were assumed to follow a Gaussian distribution in frequency.
  Using a Monte-Carlo approach, the maximum amplitude was then determined in order
to obtain comparable energy per frequency bin in the synthetic
and observed spectra.
On the other hand, except for the mode  response function, the method by \citet{Kjeldsen05} does not impose  a priori constraints concerning the modes. 
This method can then be considered to be more reliable than the method by \citet{Mosser05}.

  We compare in Fig.~\ref{velocity0} $v_{\rm HARPS}$ with the calculated mode surface velocity $v_s$ (\eq{v_s}). However, in order to have a consistent comparison, we have smoothed $v_s$  quadratically  over four radial orders. 
  We note  $\Delta v_s$ the 1-$\sigma$ error bars  associated with
$v_s$. They are derived from $\Delta \Gamma$, the 1-$\sigma$ error bars 
associated with $\Gamma$. As pointed out in Paper~I, the
uncertainty related to our knowledge of the metal abundance $Z$ for HD~49933
results in an uncertainty about the determination of  ${\cal
  P}$. However, in terms of amplitude, this uncertainty is of the order of 5~\%~; this is
negligible compared to the uncertainty that arises from  $\Delta
\Gamma$ (ranging between 25~\% to 50~\% in terms of amplitude).

The difference between computed values and observations is shown in the bottom panel of Fig.~\ref{velocity0}.
This difference must be compared with $\sigma_v$, the 1-$\sigma$
interval resulting from the errors associated with $v_s$ and this in turn associated with $v_{\rm HARPS}$,
that is $\sigma_v  \equiv \sqrt{ \Delta
  v_s^2 + \Delta v_{\rm HARPS}^2}$.
As seen in Fig.~\ref{velocity0}, except at high frequency ($\nu \gtrsim 1.9$~mHz), the theoretical $v_s$ lie well in
the 1-$\sigma_v$ domain. However, there is a clear disagreement at high frequencies where the computed mode surface velocities overestimate the observations. This disagreement is attributed to the assumptions in the theoretical model of stochastic excitation (see Sect.~\ref{Discrepancy at high frequency}). 

Assuming the 3D model with the solar abundance results in significantly larger  $v_s$.
In that case the differences between computed
 $v_s$ and the seismic constraint are in general larger than 2-$\sigma_v$. 
This shows that ignoring the metal abundance
of HD~49933 would result in a larger discrepancy between  $v_s$ and $v_{\rm HARPS}$.

\section{Amplitudes of mode in intensity}
\label{intensity}

\subsection{Derivation of mode amplitudes in intensity}
\label{calculation_intensity}

Fluctuations of the luminosity $L$ due to variations of the stellar radius
can be neglected since we are looking at high $n$ order modes; accordingly the bolometric mode intensity
fluctuations $\delta L$ are mainly due  to variations of
the effective temperature, that is:
\eqn{
{ {\delta L} \over L }    =  4  \,  {\delta T_{\rm
    eff} \over T_{\rm eff}} 
\label{dL_dTeff}
}
As in \citet{Kjeldsen95}, we now assume that $\delta T_{\rm    eff}$ is
proportional to the variation of the temperature induced by the modes at the photosphere
(i.e. at $T=T_{\rm eff}$). This assumption is discussed in
Sect.~\ref{discussion:intensity}. 
Assuming further low degree $\ell$ and \emph{adiabatic} oscillations, one can derive a relation between $\delta
T_{\rm eff}/{\rm T_{\rm eff}}$ and the radial mode velocity $v$ that is:
\eqn{
  {\delta T_{\rm  eff} \over T_{\rm eff}}  =   ( \Gamma_3 -1) \,
  \left |  {  { 1 \over
 { \omega_{\rm osc} \, \xi_{\rm r} } } \, { {{\rm d}\xi_{\rm r}} \over {{\rm d} r} }  }\right |
  \, v
\label{dTeff_ad}
}
where $\Gamma_3=
\nabla_{\rm ad}\, \Gamma_1 + 1$, $\nabla_{\rm ad}$ is the adiabatic
temperature gradient, $\Gamma_1 = \left (\derivp{\ln P_g}{\ln\rho} \right )_s$,
$\xi_{\rm r}$ the radial mode eigendisplacement, and $v$ the mode velocity \emph{at the photosphere}.
Finally, according to \eqs{dL_dTeff} and (\ref{dTeff_ad}), one has: 
\eqn{
\left ( { {\delta L} \over L } \right )   =   4 \,  \beta \, (\Gamma_3
-1)\,   \left |  {  { 1 \over
  { \omega_{\rm osc} \, \xi_{\rm r} } } \, { {{\rm d} \xi_{\rm r}} \over {{\rm d} r} } }\right 
|  \, v
\label{dL_Vad}
}
where $v$ is computed  using \eq{v_s} with $h$=0 (the photosphere), that
is:
\eqn{
v  = \sqrt{\mathcal P \over{ 2 \pi \, {\cal M}_0 \, \Gamma  } }
\label{v_0}
}
where ${\cal M}_0$ is the mode mass evaluated at the photosphere
($h$=0). 

In \eq{dL_Vad},  $\beta$ is a free parameter  introduced so that \eq{dL_Vad} 
gives, in the case of the solar p modes,  the correct maximum in
${\delta L} / L$.  Indeed, \eq{dL_Vad}  applied to the case of the solar p modes,
overestimates by $\sim 10$ times the mode amplitudes in
intensity. This important discrepancy is mainly a consequence of the
adiabatic approximation.

From the SOHO/GOLF seismic data
\citep{Baudin05}, we  derive the maximum of the solar mode (intrinsic)
surface velocity, that is  32.6~$\pm$~2.6~cm/s. Then, using $\xi_{\rm
  r}$, we  infer the
maximum of mode velocity at the photosphere,  that is  18.5~$\pm$~1.5~cm/s.
 According to \citet{Michel09}, the
maximum of the solar mode (bolometric) amplitude in intensity is equal to
2.53~$\pm$0.11~ppm. Then, by applying Eq.~(\ref{dL_Vad}) in the
case of the Sun, we derive the scaling factor $\beta=0.103~\pm~10$\,\%.
 We have checked that this calibration   depends very little on
  the choice of the chemical mixture (see also
  Sect.~\ref{discussion:intensity}).
We then adopt this value for the case of HD~49933.

\subsection{The mode intensity fluctuations measured by CoRoT}
\label{corot_intensity}

The  seismic analysis by \citet{Benomar09b} provides the apparent
amplitude $A_\ell$ of the  $\ell=$0, 1 and 2
modes and the associated error bars. 
However, the CoRoT measurements $A_\ell$ 
correspond to relative intensity fluctuations in the CoRoT
passband. Furthermore, the \emph{observed} (apparent) mode amplitudes
depend on the 
degree $\ell$. Therefore, to transform them into \emph{bolometric} and \emph{intrinsic} intensity fluctuations
\emph{normalised} to the radial modes, we
divide them by the CoRoT response function, $R_{\ell}$, derived here for
$\ell=$0, 1 and 2, 
following \citet{Michel09}.  The adopted values for $R_{\ell}$ are:
$R_0=0.90$, $R_1=1.10$, and $R_2=0.66$.
We finally derive the bolometric intensity fluctuations
normalised to the radial modes according to:
\eqn{
(\delta L/L)_{\rm
  CoRoT}  =  \sqrt{ {1 \over 3} \, \left ( \left ( A_0 \over R_0 \right )
    ^2 + \left ( A_1 \over R_1 \right )^2 + \left ( A_2 \over R_2 \right
    )^2 \right ) } \;.
}

We shall stress that the differences between the amplitudes derived by
\citet{Benomar09b} and by \citet{Appourchaux08} are smaller than
the 1-$\sigma$ error bars.  
Furthermore, these amplitudes are in agreement with those found by
\citet{Michel08}, using a different technique.

\subsection{Comparison with the CoRoT  measurements}
\label{comparison_intensity}

We compute the  mode amplitudes in terms of bolometric intensity
fluctuations, $\delta L/L$,  according  to Eqs.~(\ref{dL_Vad}) and (\ref{v_0}) (see Sect.~\ref{calculation_intensity}). 
 As for  $v_s$, the uncertainty associated with the measured mode linewidths, $\Gamma$, put
 uncertainties on the theoretical values of  $\delta L/L$. Furthermore, the uncertainty associated with the calibrated factor $\beta$ (see Sect.~\ref{calculation_intensity}) also puts an additional  uncertainty on  $\delta L/L$. 
 From here on,
 $\Delta (\delta L/L)$ will refer to the 1-$\sigma$ uncertainties
 associated with $\delta L/L$.
 Accordingly, we have $\Delta (\delta L/L) =  (\delta L/L) \, \sqrt { \left ( {1 \over 2} \, \Delta \Gamma/\Gamma \right )^2   + \left ( \Delta \beta/\beta \right )^2  } $, where $\Delta \Gamma$ (reps. $\Delta \beta$) is the 1-$\sigma$ uncertainty associated with $\Gamma$ (resp. $\beta$).

\begin{figure}[ht]        
\begin{center}
        \resizebox{\hsize}{!}{\includegraphics  {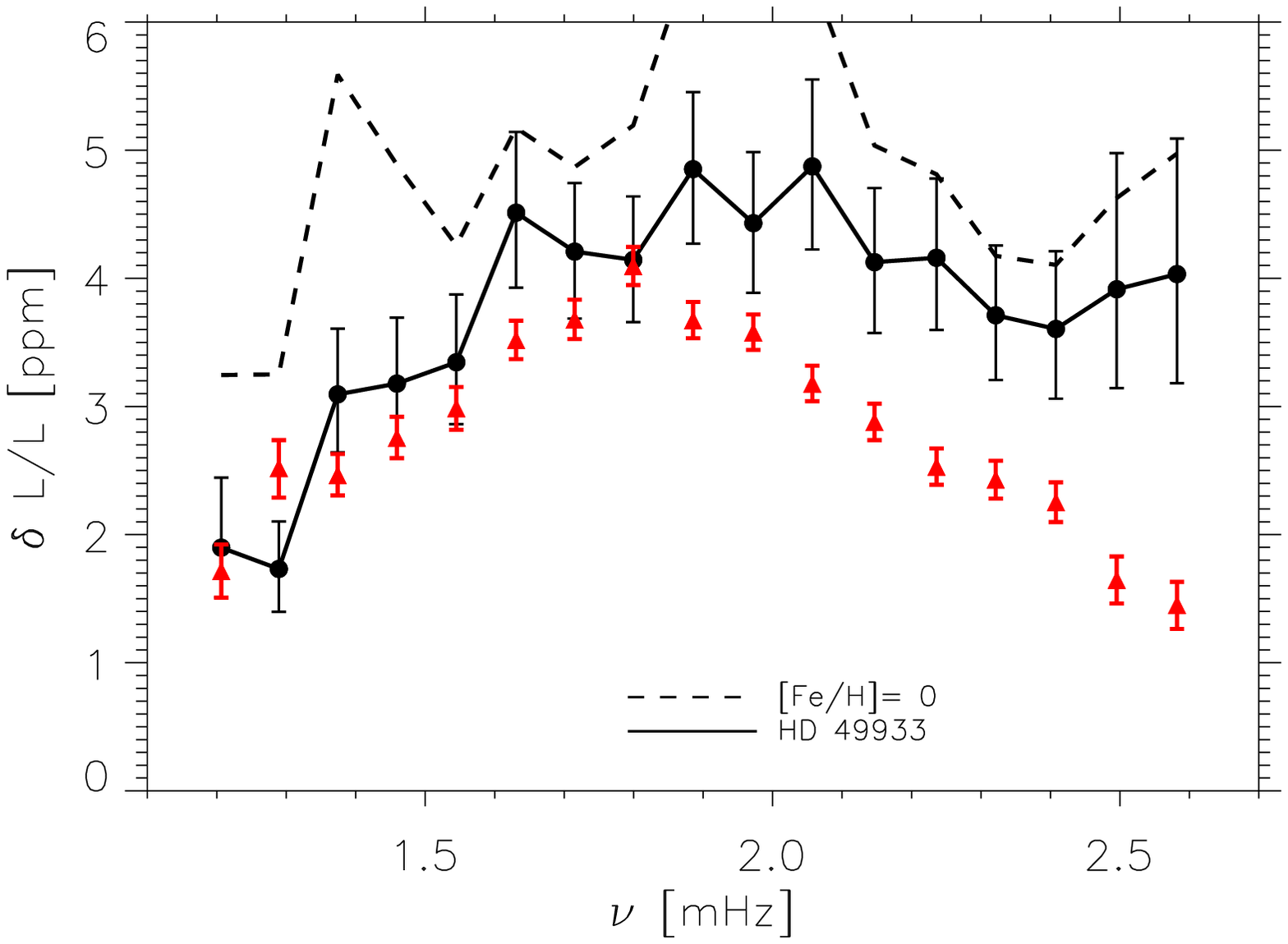}} 
         \resizebox{\hsize}{!}{\includegraphics  {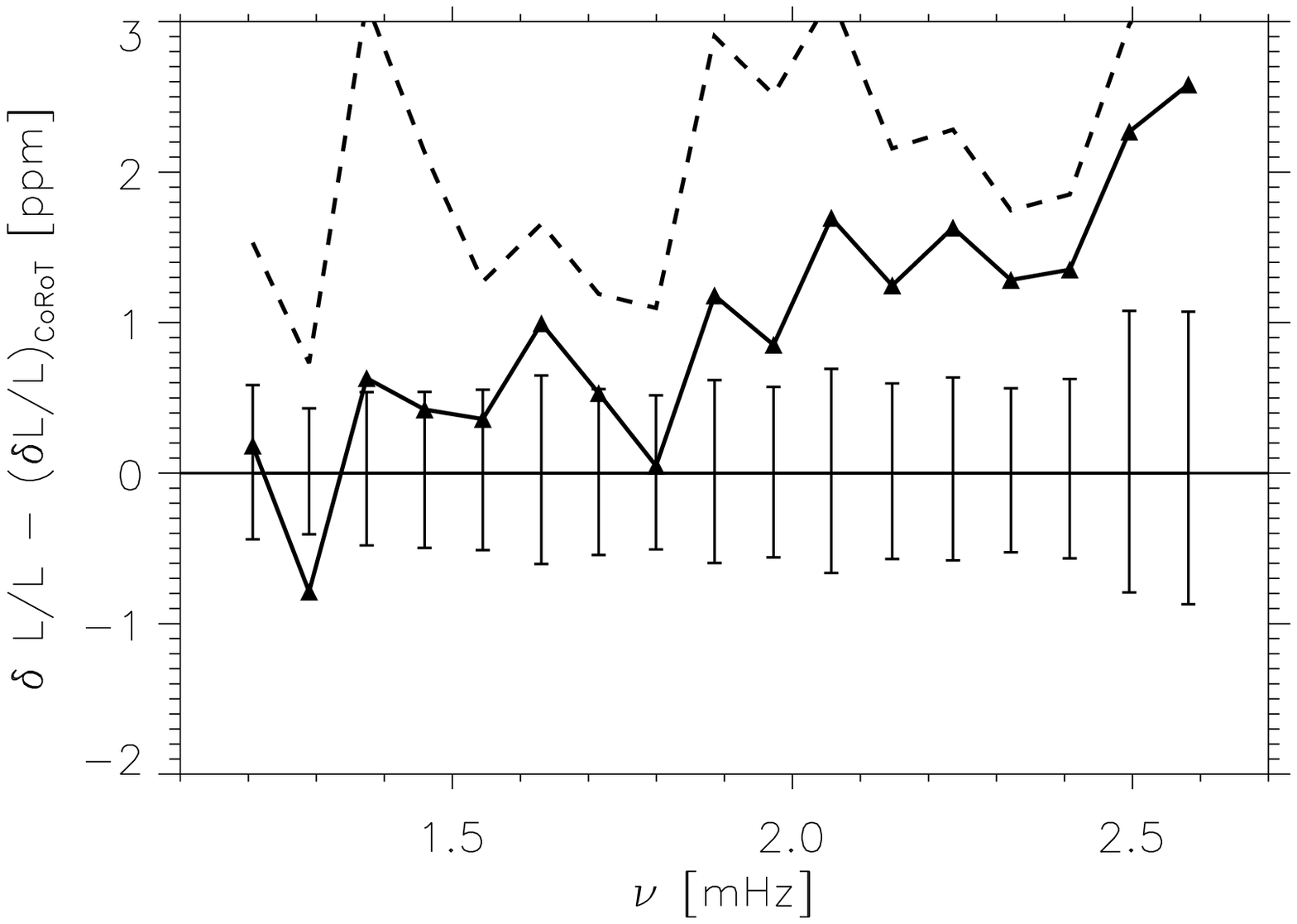}} 
       \end{center}    
        \caption{{\bf Top:} Mode bolometric amplitude in intensity as a
  function of the mode frequency ($\nu$). 
The filled circles connected by the thick solid line correspond to the
 mode amplitudes in intensity,
          $\delta L/L$, derived for HD~49933
according to Eq.~(\ref{dL_Vad}) and \eq{v_s}
where the mode surface velocity $v$ is evaluated at the photosphere.
The thick dashed solid line  corresponds to the mode amplitude in intensity 
associated with the model with [Fe/H]$=0$. 
The red triangles and associated error bars
  correspond to the mode amplitudes in intensity, $(\delta L/L)_{\rm CoRoT}$, obtained by
 from the CoRoT data \citep{Benomar09b}. These measurements have been translated into bolometric
  amplitudes following \citet{Michel09}.
 {\bf Bottom:} Same as top for the difference between $\delta L/L$ and
       $(\delta L/L)_{\rm CoRoT}$.  The 1-$\sigma$ error bars
 correspond here to $\sqrt{ 
   a  ^2 +  b  ^2 }$ where $a \equiv \Delta (\delta L/L) $ and $b
 \equiv \Delta(\delta L/L)_{\rm CoRoT}$ (see text).} 
        \label{intensity0}
        \end{figure}

Figure~\ref{intensity0} compares, as a function of the mode
frequency,  $\delta L/L$  to the CoRoT measurements: $(\delta L/L)_{\rm
  CoRoT}$.   
The difference between our calculations and the observations is shown in the
bottom panel. As for the velocity, this difference must be compared with
$\sigma_L$, the 1-$\sigma$ interval resulting from the association of the
1-$\sigma$ error bars $\Delta (\delta L/L)$ and the 1-$\sigma$ error,
$\Delta(\delta L/L)_{\rm CoRoT}$, associated 
with the CoRoT measurements.  Accordingly, we have $\sigma_L
\equiv \sqrt{  a  ^2 +  b  ^2 } $ where $a \equiv \Delta (\delta L/L) $ and $b
 \equiv \Delta(\delta L/L)_{\rm CoRoT}$.

As seen in Fig.~\ref{intensity0}, below $\nu \lesssim 1.9$~mHz, 
values of $\delta L/L$ are  within approximately 1-$\sigma_L$ in agreement with $(\delta
L/L)_{\rm CoRoT}$. 
However, above $\nu \sim 1.9$~mHz, the differences between $\delta L/L$ and $(\delta
L/L)_{\rm CoRoT}$ exceed 2-$\sigma_L$.

Assuming a solar abundance ([Fe/H]$=0$) results in a clear overestimation of $\Delta(\delta L/L)_{\rm CoRoT}$. 
Furthermore, calculations which assume the \citet{GN93} chemical mixture
result in mode amplitudes larger by $\sim$ 15\,\%. 

 Both in terms of intensity and velocity, differences between the
calculated mode amplitudes and those derived from the observations
(CoRoT and HARPS) are approximately within the 1-$\sigma$ domain below
$\nu \sim 1.9$~mHz. This then validates  the intensity-velocity relation given by \eq{dL_Vad} at the level of the current
seismic precision .

The maximum $(\delta L/L)$ peaks at $\nu_{\rm max}
\simeq 1.9$~mHz and the maximum of $v_s$ at $\nu_{\rm max}
\simeq 1.8$~mHz. By comparison, $(\delta L/L)_{\rm CoRoT}$ peaks at $\nu_{\rm
  max} \simeq 1.8$~mHz and $v_{\rm HARPS}$ peaks at  $\nu_{\rm
  max} \simeq 1.7$~mHz.  
The difference in  $\nu_{\rm  max}$ between the observations (CoRoT and HARPS) and the model can be partially a
consequence of the clear tendency at high frequency toward
over-estimated amplitudes compared to the observations.

\section{Discussion}
\label{Discussion}

\subsection{Uncertainties in the knowledge of the fundamental parameters of HD~49933}

Uncertainties in the knowledge of $T_{\rm
  eff}$ and $\log~g$ place uncertainties on the theoretical values of
${\cal P}$ and hence on the mode amplitudes ($v_s$ and $\delta L/L$). 
However, estimating these  uncertainties would require the
consideration of  3D models with
a $T_{\rm eff}$ and a $\log g$ that depart  more than 1-$\sigma$ from the
 values adopted in our modeling, i.e.  $T_{\rm  eff} = 6750~$K and
 $\log~g = 4.25$. This is beyond the scope of our efforts since such 3D models are not
 yet available. 

 \subsection{Influence of the mode mass}

As discussed in details in \citet{Samadi08}, the computed mode surface
velocities $v_s$ significantly depend on the choice of the height $h$
in the atmosphere where the mode masses are evaluated.   
According to \citet{Samadi08}, seismic measurements performed with the
HARPS spectrograph reflect conditions slightly below the formation depth of the
K line. Accordingly, we have evaluated  by default
  the mode masses at the optical depth where the K line is
  expected to be formed (i.e. $\tau_{\rm 500~nm}
\simeq 0.013$), which corresponds, for our 3D models, to a height of about
350~km above the photosphere. 
We can evaluate how sensitive we are to  the choice of $h$. Indeed, evaluating the mode mass at the photosphere
results in values of $v_s$ which are about 15\,\% lower and hence would reduce the discrepancy with
the HARPS observations. On the other hand, evaluating the mode mass one
pressure scale height ($\sim$~300~km at the photosphere) above $h =
350$~km results in an increase of $v_s$ of about 10\,\%.
A more rigorous approach to derive the different heights in the
atmosphere where the measurements are sensitive would require a
dedicated modeling \citep[see a discussion in][]{Samadi08}.

\subsection{The intensity-velocity relation}
\label{discussion:intensity}

\bigskip
 {\noindent \it Sensitivity to the location:}    \\
The derivation of \eq{dTeff_ad} (or equivalently \eq{dL_Vad}) is based on the  assumption 
 that  $\delta T_{\rm eff} \propto \left . \delta T \right |_{T=T_{\rm
     eff}}$ (see Sect.~\ref{calculation_intensity}). 
This is  quite a arbitrary simplification. In order to check how
sensitive our results are to this assumption, we have
computed \eq{dL_Vad} and \eq{v_0} for two different positions
in the atmosphere. The first position, $h = h_{1}$, is chosen one pressure 
scale height ($\simeq 300$~km) above the photosphere, which corresponds to an
optical depth of $\tau \sim 0.02$. The second position, $h = h_2$, is chosen one
pressure scale height beneath the photosphere, that is around $\tau
\sim 200$.
For both positions, the mode amplitudes with frequencies below $\sim 1.9$~mHz are almost
unchanged. Concerning  the amplitudes of modes with frequencies above  $\sim
1.9$~mHz, they are increased by up to $\sim$~20~\%  when $h=h_1$ and are in turn almost
unchanged when $h=h_2$.  
Since the fluctuations of $L$ induced by the
oscillations are mostly due to temperature changes that occur around
an optical depth of the order of the unit,  we can conclude that our
calculations are almost insensitive to the choice of the layer in the
visible atmosphere where $\delta T$ is evaluated.

\bigskip
{\noindent \it Non-adiabatic effects:}   \\
  The modes are measured at the surface of the star where non-adiabatic
  interactions between the modes and convection as well as radiative
  losses of the modes are important. Assuming  
  \eq{dTeff_ad} is then a crude approximation. In fact, it is clearly non-valid in the case of the Sun since it results 
  in  a severe over-estimation of the solar mode amplitudes in 
  intensity (see Sect.~\ref{calculation_intensity}). 
  Avoiding this approximation requires non-adiabatic eigenfunctions
  computed with a  time-dependent
  convection model. However, such models
  \citep[e.g.][]{Grigahcene05,Balmforth92a} are  subject to large 
  uncertainties, and there is currently no consensus about the 
   non-adiabatic mechanisms that play a significant role \citep[see
     e.g. the recent review by][]{Houdek08}.   
   For instance, parameters are usually introduced in the theories so
   that  they cannot be used in a predictive way.  

   In the present study, we adopt by default the adiabatic
   approximation and introduce in  \eq{dL_Vad} the parameter $\beta$ calibrated
   with helioseismic data. 
   We show here that despite the deficiency of the quasi-adiabatic
   approximation, it nevertheless provides the correct scaling, at
   least at low frequency and at the level of the present seismic precisions. 

   As an alternative approach, comparing the spectrum obtained from
   the 3D models in intensity with that obtained in velocity can provide
   valuable information concerning the intensity-velocity relation, in
   particular concerning the departure from the adiabatic
   approximation  and the sensitivity to the surface metal abundance.
   We have started to carry out such a study. For the velocity, the (few) acoustic modes
   trapped in the simulated boxes can  be extracted and their
   properties measured. 
   But this was impossible to do for the intensity with the simulations at our
   disposal because the computed spectrum for the  intensity is dominated by the
   granulation background. As a consequence it is
   not possible to extract the mode amplitudes
   in intensity with sufficient accuracy . 
   A comparison between the spectra obtained from the 3D models
   requires a much longer time series (work in progress). 

\bigskip
{\noindent \it Sensitivity to the metal abundance:}  \\
   We have shown in this study how  the mode amplitudes
   in the velocity are sensitive to the surface metal abundance. An open question is how
   sensitive is  the intensity-velocity relation in general to the
   metal abundance? 
   A theoretical answer to this question would require a realistic and
   validated non-adiabatic treatment. 
   The pure numerical approach mentionned above can also in
   principle provide  some answers to this question. However, as discussed above,
   this approach is not applicable with the time series at our disposal.   
   Concerning the quasi-adiabatic relation of \eq{dL_Vad}: a change
   of the metal abundance has a direct effect on $\Gamma_3$ and an
   indirect effect on the properties of the (radial)
   eigen-displacement $\xi_{\rm r}$. However, the comparison between
   the metal-poor 3D model (S1) and the 3D model with the solar
   abundances (S0)  shows that --~ at a fixed frequency $\omega_{\rm
     osc}$ ~-- the ratio $(\delta L/L) / v$, which is equal to  $
   4 \, \beta \, (\Gamma_3-1) \,  ({\rm d} \xi_{\rm r}/ {\rm d} r)/ (\xi_{\rm r} \,
   \omega_{\rm osc})$,  is almost unchanged between S0 and S1 (the
   differences are less than $\sim$~1\,\%). 
   In conclusion, the quasi-adiabatic relation
   of \eq{dL_Vad} depends very weakly on the surface metal
   abundance. Accordingly, the choice of the
   solar chemical mixture has  a negligible impact on the value
   of the calibration factor $\beta$.

\subsection{The solar case}

 As seen in this study, the surface metal abundance has a pronounced
 effect on the mode excitation rates. One may then wonder about the
 previous validation of the theoretical model of stochastic excitation
 in the case of the Sun \citep{Kevin06b,Samadi08b}. 
 Indeed, this validation was carried out with the use of a
 solar 3D model based  on an ''old'' solar chemical mixture
 \citep[namely those proposed by][]{Anders89} while
the ''new'' chemical mixture by \citet{Asplund05} is characterized
 by a significantly lower metal abundance.  

 In order to adress this issue, we  have first considered two global 1D solar
   models. One model has an ''old'' solar abundance \citep[][model M$_{\rm
     old}$ hereafter]{GN93}  while the second one 
   has  the ''new'' abundances   \citep[][model M$_{\rm
     new}$ hereafter] {Asplund05}.
      At the surface where the excitation occurs, the density of the
   solar model M$_{\rm new}$ is only $\sim$~5 \% lower
   compared to the model M$_{\rm old}$.
   According to the arguments developed in Paper I, this difference in the density must
   imply a difference in the convective velocities ($\tilde u$) of the
   order of $\sim (\rho_{\rm old}/\rho_{\rm new})^{1/3}$, where
     $\rho_{\rm old}$ (resp. $\rho_{\rm new}$) is the surface density 
     associated with M$_{\rm old}$  (resp. M$_{\rm new}$).
     Accordingly, $\tilde u$ is expected to be $\sim$ 1.7 
     \% higher for  M$_{\rm new}$ compared to M$_{\rm old}$. 

   The next question is what is the change in the solar mode
   excitation rates induced by the above difference in  $\tilde u$? 
   We have computed the solar mode excitation rates exactly in the
   same manner as for HD 49333 by using a solar 3D simulations based on
   the ''old'' abundances.
 We obtained a rather good agreement with the different helioseismic
 data \citep[see the result in][]{Samadi08b}. 
 To derive the solar mode excitations expected with the ''new'' solar
   abudance, we have proceeded in a similar way as the one done in
   Paper I: we have increased the convective velocity  $\tilde
   u$ derived from the solar 3D model by 2 \%  while keeping the kinetic flux
   constant (see details in Paper I). 
    This increase of $\sim $ 2 \% of $\tilde u$ results in an increase
    of $\sim$~10 \% of the mode excitation rates. This increase is 
   significantly lower than the current uncertainties associated with
   the different helioseismic data \citep[][]{Baudin05,Samadi08b}.  
%

\subsection{Discrepancy at high frequency}
\label{Discrepancy at high frequency}

The discrepancy betwen  theoretical calculations and  observations
is particularly pronounced at high frequency.
This discrepancy  may be attributed to
a canceling between the  entropy  and the Reynolds stress contributions
(see Sect.~\ref{canceling_entropy_reynolds})
or the "scale length separation" assumption  (see
Sect.~\ref{scale length separation}).

\subsubsection{Canceling between the entropy  and the Reynolds stress contributions}
\label{canceling_entropy_reynolds}

\begin{figure}[ht]
        \begin{center}
        \resizebox{\hsize}{!}{\includegraphics  {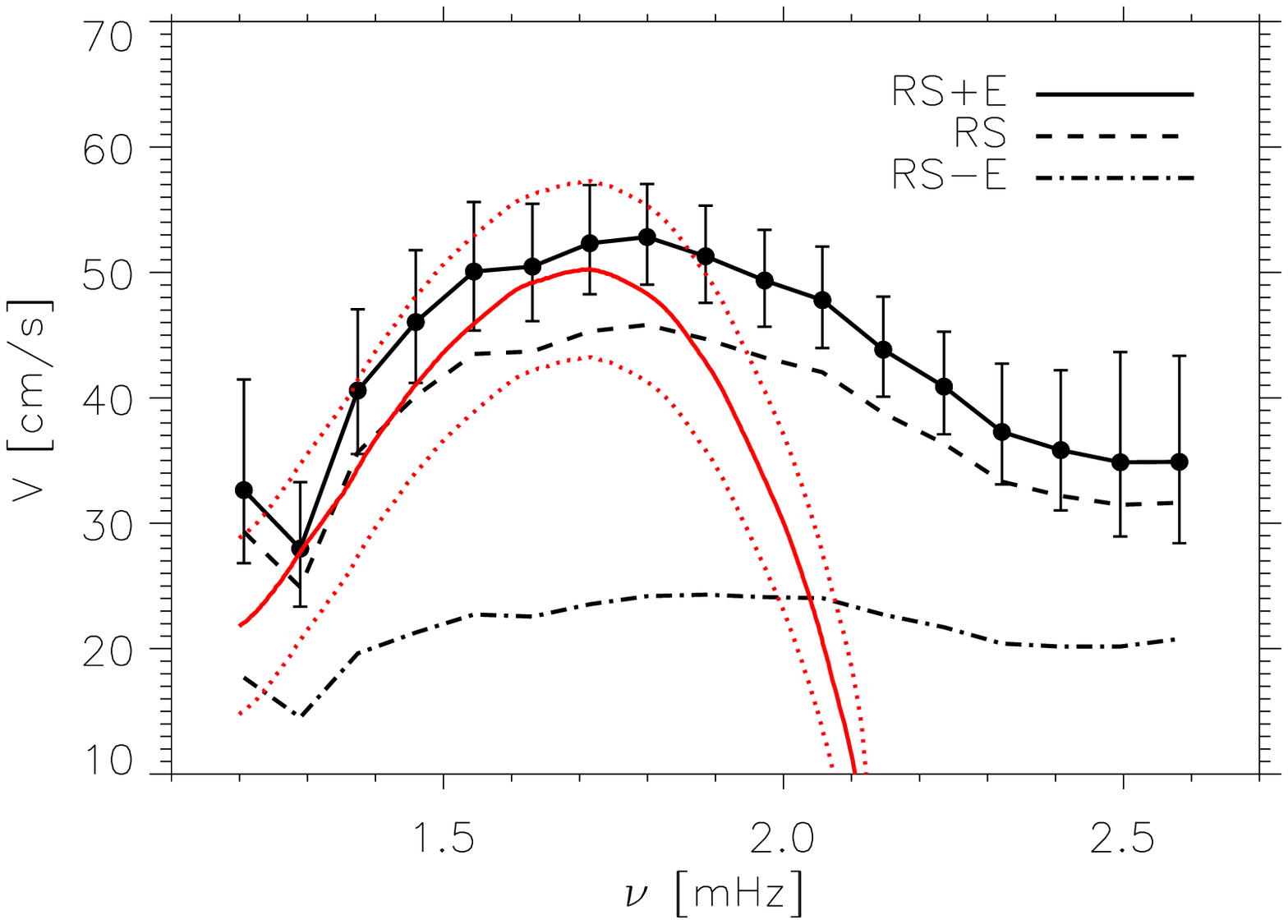}} 
        \resizebox{\hsize}{!}{\includegraphics  {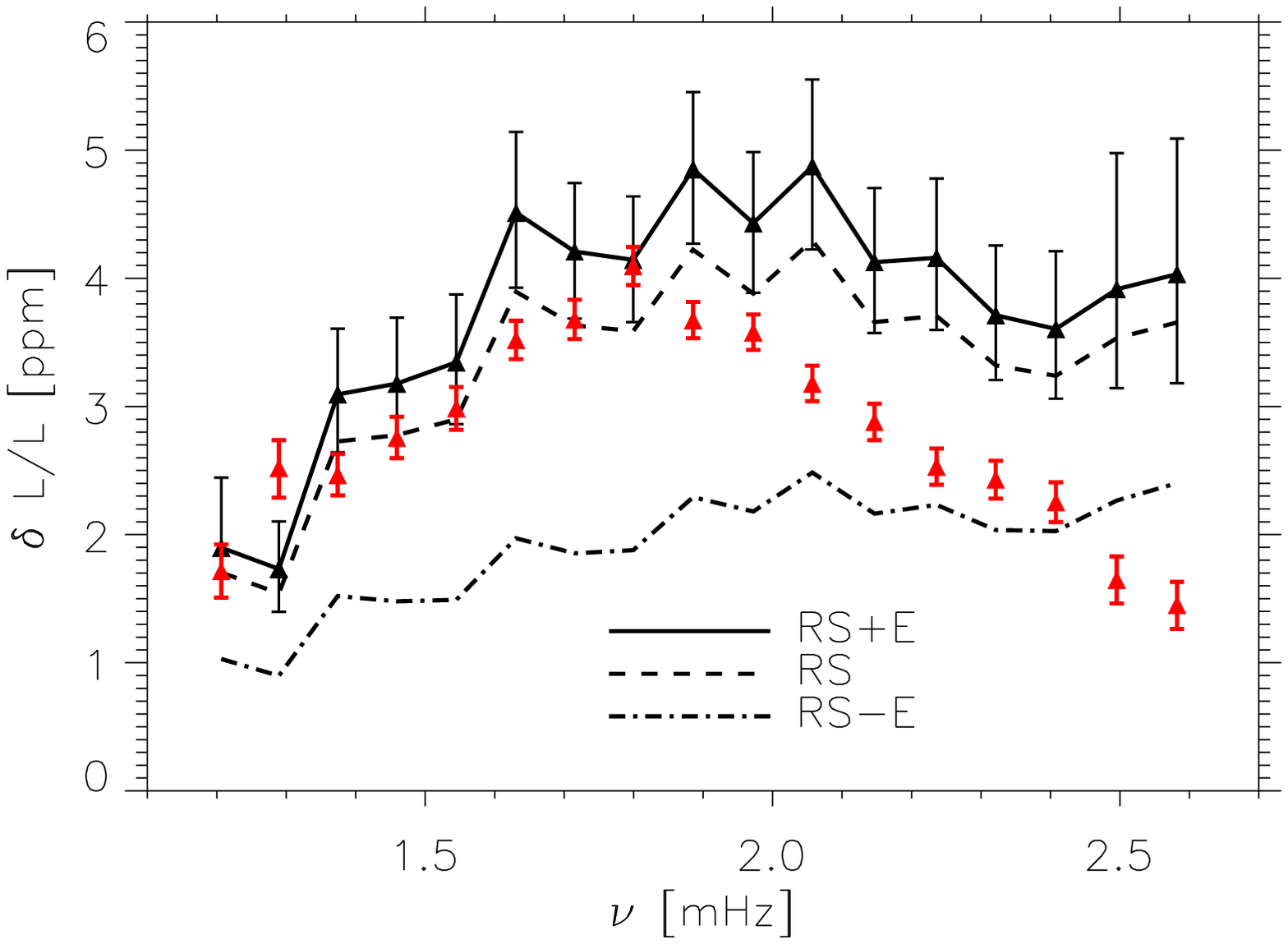}} 
    \end{center}    
        \caption{{\bf Top:} Same as Fig.~\ref{velocity0}.  
The thin dashed line corresponds to a calculation that
takes only the contribution of the Reynolds stress into account. The
dot-dashed line corresponds to a calculation in which we have assumed
that the contribution of the Reynolds stress interferes totally with that
of the entropy fluctuations (see text). The thick
  solid line has the same meaning as in Fig~\ref{velocity0}. {\bf
    Bottom:} Same as top for $\delta L/L$. The triangles and associated error bars have the same meaning as in Fig.~\ref{intensity0}}
        \label{entropy}
        \end{figure}

The relative contribution of the entropy fluctuations to the
excitation is found  to be about 30\,\% of the total excitation. 
This is two times larger than in the case of the Sun ($\sim $15\,\%).  
This can be explained by the fact HD~49933 is significantly hotter than the
Sun and, as pointed-out by \citet{Samadi07a}, the larger $(L/M) \propto
T_{\rm eff}^4/g$, the more important the relative contribution of the
entropy.  
Although more important than in the Sun, the contribution of the entropy fluctuations
 remains relatively smaller than the uncertainties associated with the
 current seismic data. 
This is illustrated in Fig.~\ref{entropy}: the difference between
theoretical mode amplitudes which take into account only the Reynolds stress contribution ($C_R^2$, see
Eq.~(3) of Paper~I) and those that include both
contributions (entropy and Reynolds stress) is lower than $\sigma_v$. 
In terms of amplitudes, the entropy fluctuations contribute only
$\sim$ 15\,\% of the global amplitude. This is significantly smaller
than the uncertainties associated with the current seismic
measurements. Seismic data of a better quality are then needed to
constrain the entropy contribution and its possible canceling with the
Reynolds stress. 

Numerical simulations show some cancellation between the
entropy source term and the one due to the Reynolds stress \citep{Stein04}. 
However, in the present theoretical model of stochastic excitation, 
the cross terms between the entropy fluctuations and the Reynolds
stresses vanish \citep[see][]{Samadi00I}. This is a consequence of
the different assumptions concerning the entropy fluctuations 
 \citep[see][see also the recent discussion in \citet{Samadi08b}]{Samadi00I}. 
Accordingly, the entropy source
term is included as a source independent from the Reynolds stress
contribution. 
As suggested by \citet{Houdek06}, a partial
canceling between the entropy fluctuations and the Reynolds stress can
 decrease the mode amplitudes of F-type stars and reduce the
discrepancy between the theoretical calculations and the observations.

There is currently no theoretical
description of these interferences. In order to have an upper limit
  of the  interferences, we assume that both
contributions \emph{locally} and \emph{fully} interfer. 
This assumption leads to the computation of 
 the excitation rates per unit mass as:
\begin{equation}
{ {d {\cal P}}\over {d m}} = \left ( { {d {\cal P}}\over {d m}} \right
)_{RS} + \left (  { {d {\cal
      P}}\over {d m}}  \right )_{E} - 2 \sqrt{  \left (  { {d {\cal P}}\over {d
        m}} \right )_{RS} \,  \left (  { {d {\cal
      P}}\over {d m}} \right )_{E}  }
\label{cancel}
\end{equation}
where $\left ( { {d {\cal P}} /  {d m}} \right
)_{RS} $ and  $ \left (  { {d {\cal  P}} / {d m}} \right )_{E} $ are the
contributions per unit mass of the Reynolds stress and entropy
respectively. 
The result is presented in
Fig.~\ref{entropy} in terms of velocity (top pannel) and in terms of
intensity (bottom pannel). The mode amplitudes are decreased by up to
$\sim 55$\,\%. In that case, $(\delta L/L)_{\rm CoRoT}$ is
systematically under-estimated. Obviously, a partial canceling between the entropy
contribution and the Reynolds stress would result in a smaller
decrease. 

We have assumed here that the cancellation between the two terms is independent
of the mode frequency (see \eq{cancel}). 
However, according to \citet{Stein04}, the level of the
  cancellation depends on the frequency (see their Fig.~8).
  In particular, for F-type stars, the cancellation is expected to be
  more important around and  above the peak frequency. 

As a conclusion, the existence of a partial canceling between the entropy fluctuations and the Reynolds stress can
decrease the mode amplitude and could improve  the agreement with the seismic
observations at high frequency. 
However, there is currently no theoretical modeling of the
interference between theses two terms. Further theoretical
developements are required.

\subsubsection{The "scale length separation" assumption }
\label{scale length separation}

 The "scale length separation" assumption \citep[see the  
  review by][]{Samadi08b} consists of the assumption that 
the eddies contributing effectively to the driving  have a characteristic length scale smaller than the mode wavelength.
This assumption is justified for a low Mach number ($M_t$).
However, this approximation is less valid in the super-adiabatic
region where $M_t$ reaches a maximum (for
the Sun $M_t$ is up to 0.3) and accordingly affects the high-frequency modes more. 
This approximation is then expected to be even more questionable for stars hotter
than the Sun, since $M_t$ increases with $T_{\rm eff}$. 
This spatial separation can be avoided,  however if the kinetic energy spectrum associated with the turbulent elements
($E(k)$) is properly coupled with the spatial dependence of the modes
(work in progress).  In that case, we
expect  a more rapid decrease of the driving efficiency with
increasing frequency than in the present formalism where the spatial
dependence of the modes is  totally decoupled from $E(k)$
(i.e. ''scale length separation'').

\section{Conclusion}
\label{Conclusion}

From the mode linewidths measured by CoRoT  and
theoretical mode excitation rates derived for HD~49933, we have derived
the expected mode surface velocities $v_s$ which we have compared with 
$v_{\rm HARPS}$, the mode velocity spectrum derived from the seismic observations obtained with the HARPS spectrograph
\citep{Mosser05}. 
Except at high frequency ($\nu \gtrsim$~1.9~mHz), the agreement between  computed $v_s$ and
$v_{\rm  HARPS}$ is within the 1-$\sigma$ domain associated with the
seismic data from the HARPS spectrograph. 
However,  there is a clear tendency to overestimate
 $v_{\rm  HARPS}$ above $\nu~\sim$~1.9~mHz,.

 Using a \emph{calibrated} quasi-adiabatic approximation to relate the mode velocity to the mode
amplitude in intensity (Eq.~\ref{dL_Vad}), we have derived for the case of HD~49933 the
expected mode amplitudes in intensity. 
 Computed mode intensity fluctuations, $\delta L/L$, 
are within 1-$\sigma$  in agreement with the seismic
constraints derived from the CoRoT data \citep{Benomar09b}.  
However, as for the velocity, there is a clear tendency at high frequency ($\nu \gtrsim$~1.9~mHz) towards
over-estimated $\delta L/L$ compared to the CoRoT observations.

Calculations that assume a solar surface metal abundance result, both in velocity and in
intensity, in amplitudes larger by $\sim$~35\,\% around the peak
frequency ($\nu_{\rm max} \simeq$ 1.8~mHz) and by up to a factor of two at
lower frequency. It follows that, ignoring the current surface metal abundance of
the star results in a more severe over-estimation of the 
computed amplitudes compared with observations. 
This illustrates the importance of taking the surface metal abundance of
the solar-like pulsators into account when modeling the mode driving.
In addition, we point out that the \citet{GN93} solar chemical mixture
 results in mode amplitudes larger by about 15\,\% with respect to
 calculations that assume the ''new'' solar abundance by
 \citet{Asplund05}. However, this increase remains significantly
 smaller than the  uncertainties associated with current seismic measurements. 

Since both mode amplitudes in terms of surface
velocity and intensity are available for this star, it was 
possible to test the validity of the calibrated quasi-adiabatic relation (\eq{dL_Vad}).  
Our comparison  shows that this relation   provides the correct scaling, at least at the level of the present seismic precisions,  .

 Both in terms of surface velocity and of intensity, the
differences between predicted and observed mode amplitudes  are within
the 1-$\sigma$ uncertainty domain, except at high frequency. 
This result then validates for low frequency modes the basic underlying physical assumptions included in the theoretical
model of stochastic excitation  for a star significantly different in
effective temperature, surface gravity, turbulent Mach number ($M_t$) and metallicity compared to the
Sun or $\alpha$~Cen~A. 

As discussed in Sect.~\ref{Discussion}, 
the clear discrepancy  between predicted
and observed mode amplitudes  seen at high frequency may have two
possible origins:
First, a canceling between the entropy contribution and the Reynolds
stress is expected to occur and to be important around and above the
 frequency of the maximum of the mode excitation rates (see Sect.~\ref{canceling_entropy_reynolds}). 
Second, the assumption called the ``scale length separation'' 
\citep{Samadi08b}
may also result in an over-estimation of the mode amplitudes at high
frequency (see Sect.~\ref{scale length separation}). 
These issues will be investigated in a forthcoming paper.

\begin{acknowledgements}
The CoRoT space mission, launched on
December 27 2006, has been developed and is operated by CNES,
with the contribution of Austria, Belgium, Brasil, ESA, Germany
and Spain.\\
We are grateful to the referee for his pertinent comments.
We are indebted to J. Leibacher for his careful reading of the
manuscript. 
K.B. acknowledged financial support from Li\`ege University through the Subside F\'ed\'eral pour la Recherche 2009. 
\end{acknowledgements}
\bibliographystyle{aa}

\end{document}